\documentclass{article}

\usepackage{graphicx}
\usepackage{pdfpages}

\begin{document}

\includepdf[pages=-]{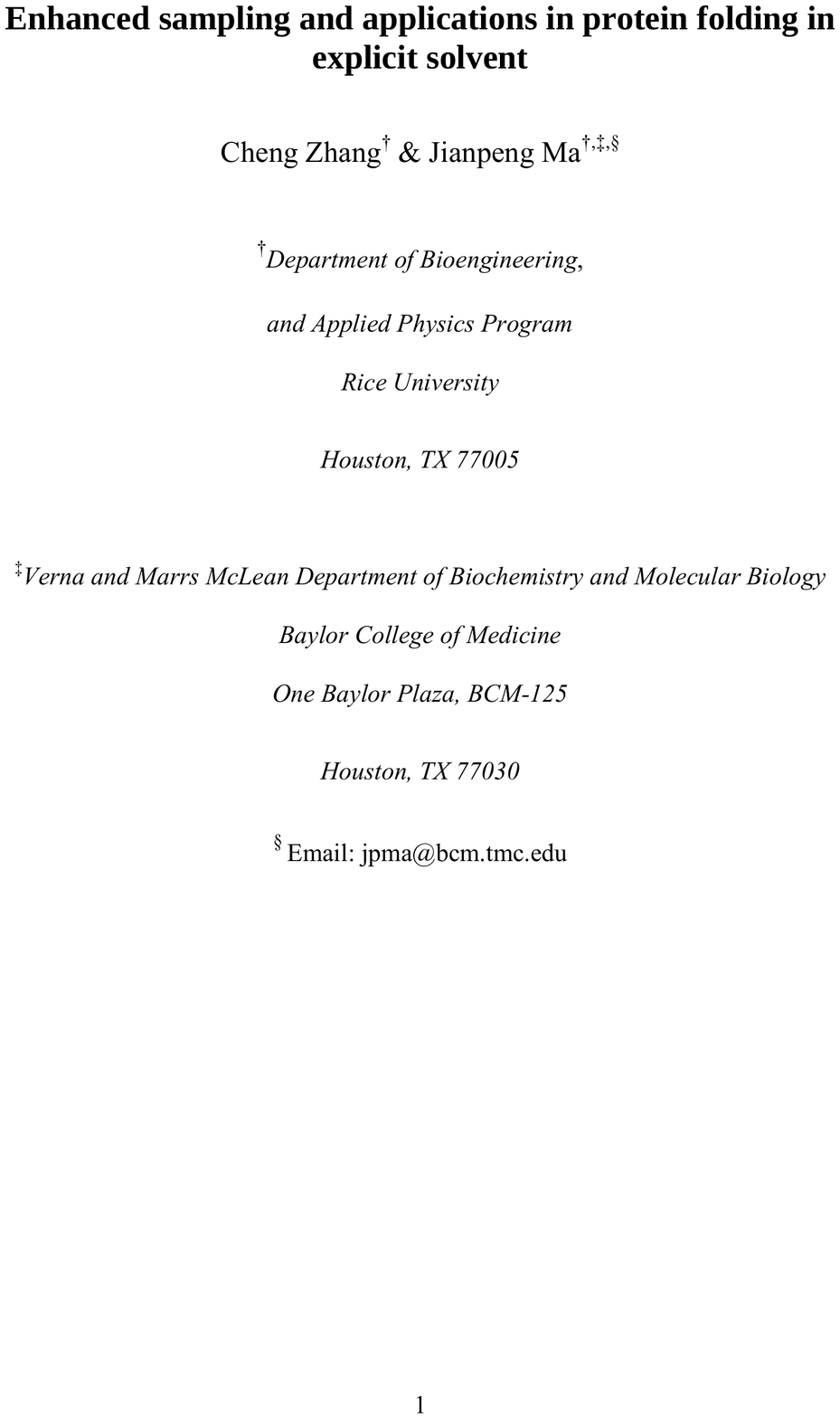}

\begin{figure}
  \begin{minipage}{ 0.95 \linewidth}
    \begin{center}
        \includegraphics[angle=0,width=\linewidth]{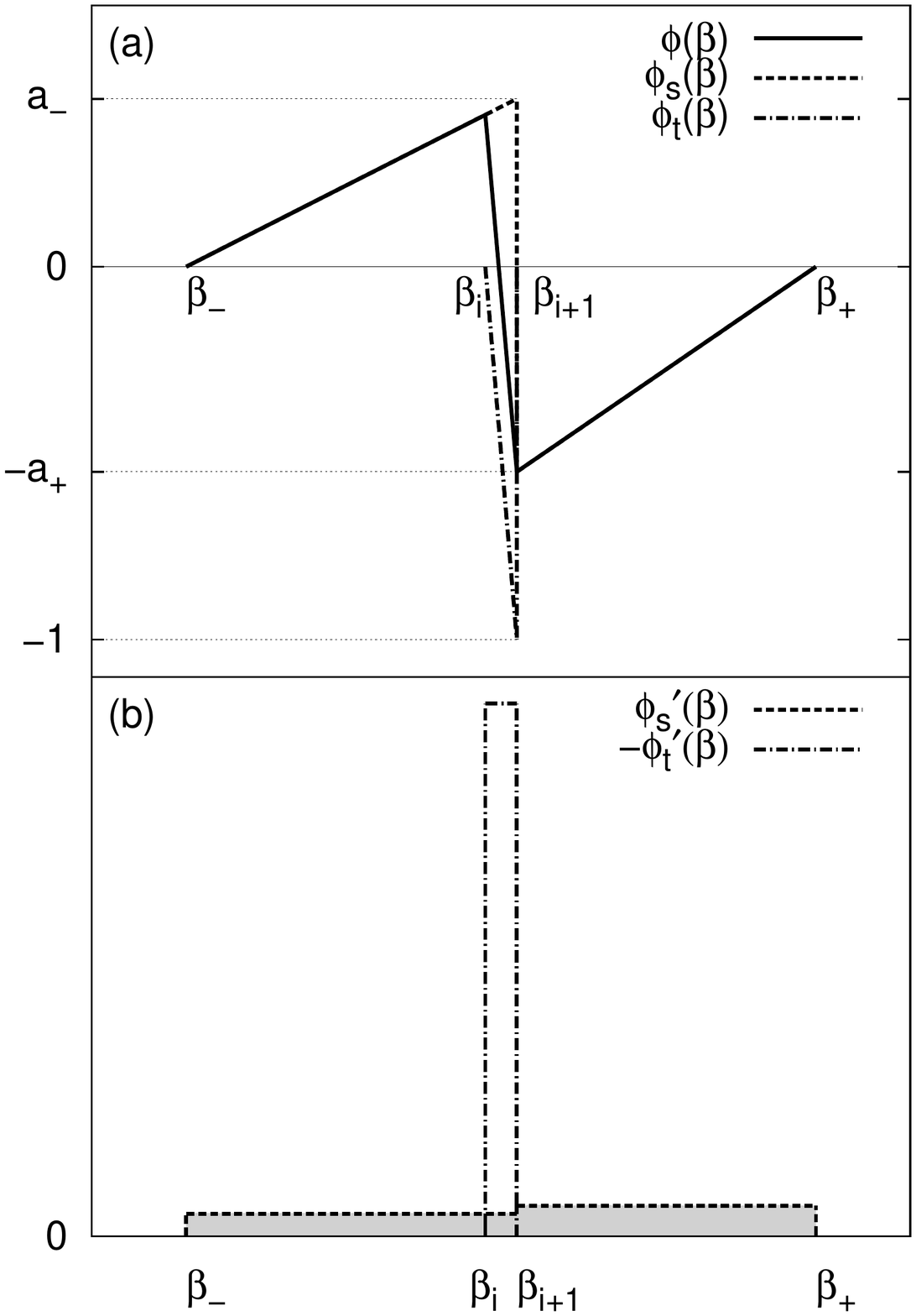}
    \end{center}
  \end{minipage}%
  \caption{\label{fig:phi} }
\end{figure}


\begin{figure}
  \begin{minipage}{ 0.95 \linewidth}
    \begin{center}
        \includegraphics[angle=0,width=\linewidth]{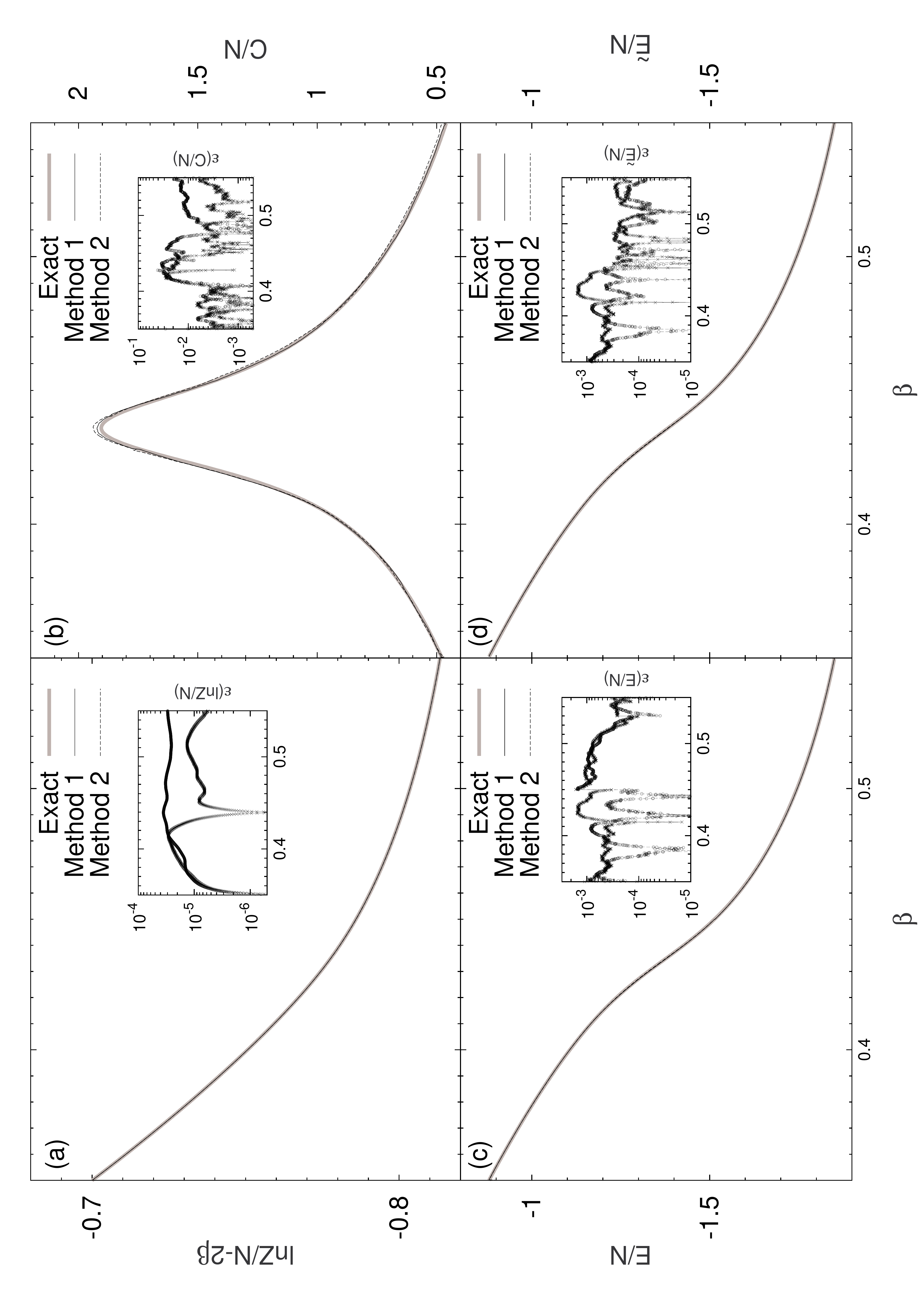}
    \end{center}
  \end{minipage}%
  \caption{\label{fig:isingall} }
\end{figure}

\begin{figure}
  \begin{minipage}{ 1.0 \linewidth}
    \begin{center}
        \includegraphics[angle=-90,width=\linewidth]{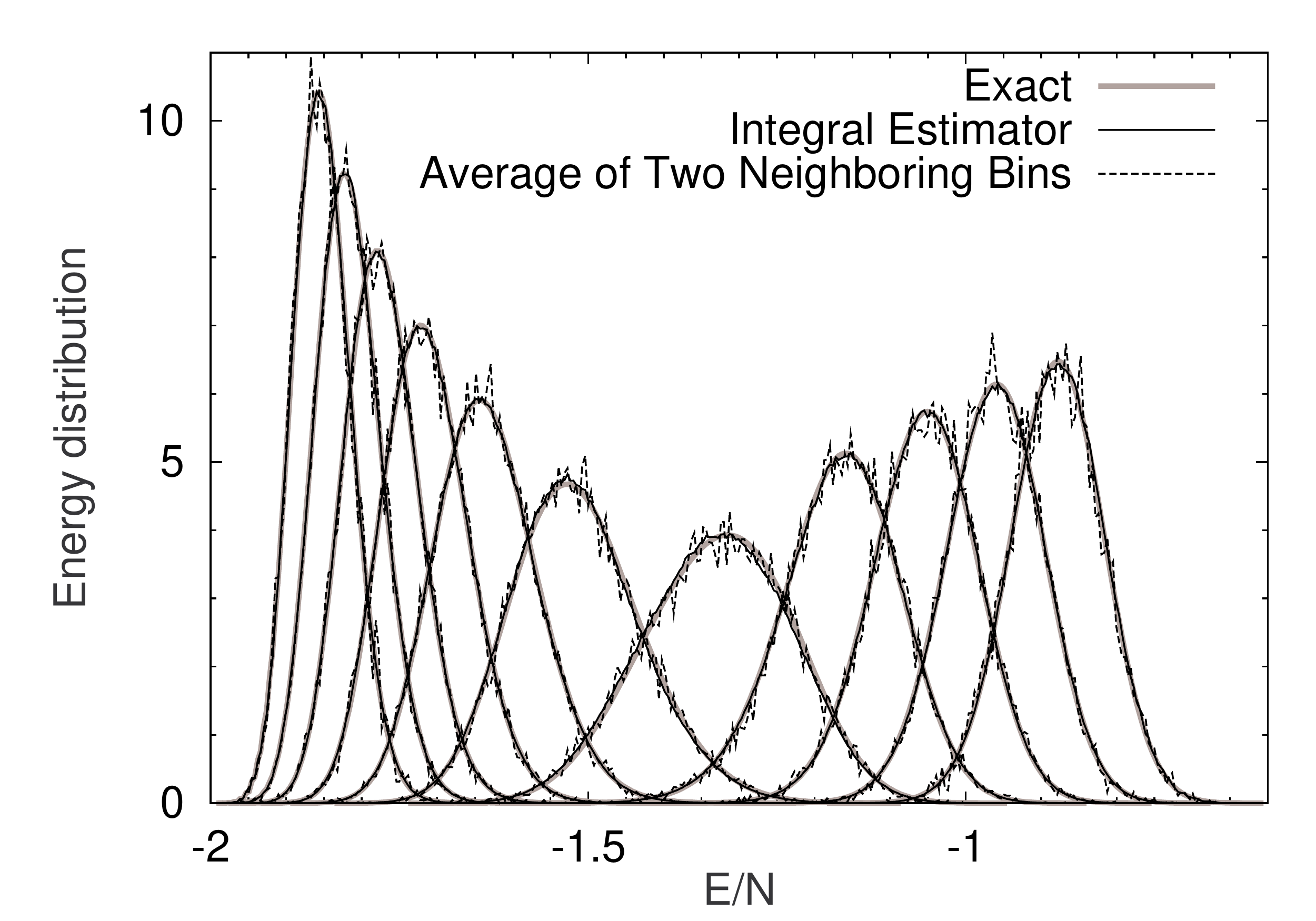}
    \end{center}
  \end{minipage}%
  \caption{\label{fig:isinghist} }
\end{figure}


\begin{figure}[ht]
  \begin{minipage}{ 1.0 \linewidth}
    \begin{center}
        \includegraphics[width=\linewidth]{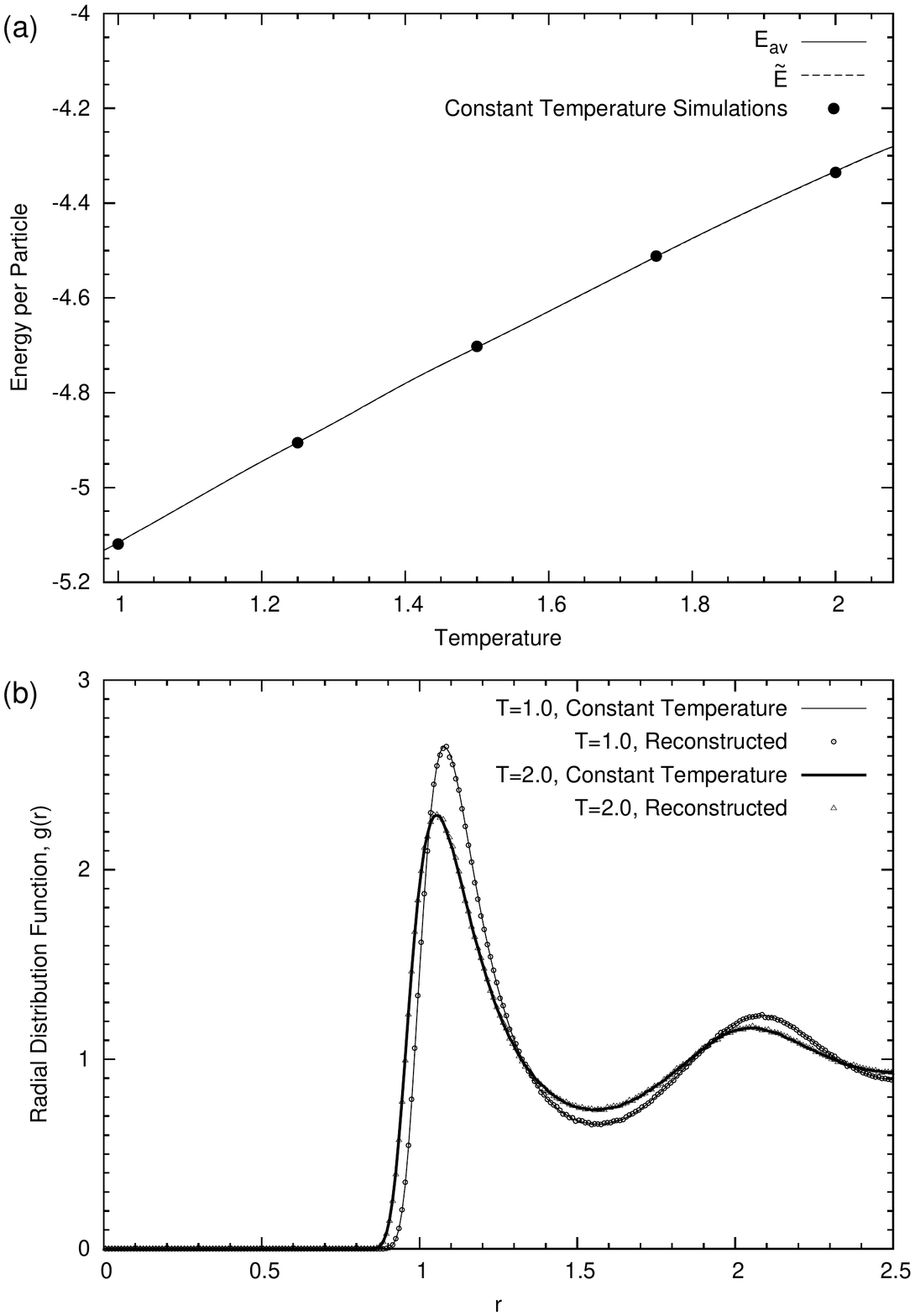}
    \end{center}
  \end{minipage}%
  \caption{\label{fig:lj} }
\end{figure}


\begin{figure}[ht]
  \begin{minipage}{ 1.0 \linewidth}
    \begin{center}
        \includegraphics[angle=90,width=\linewidth]{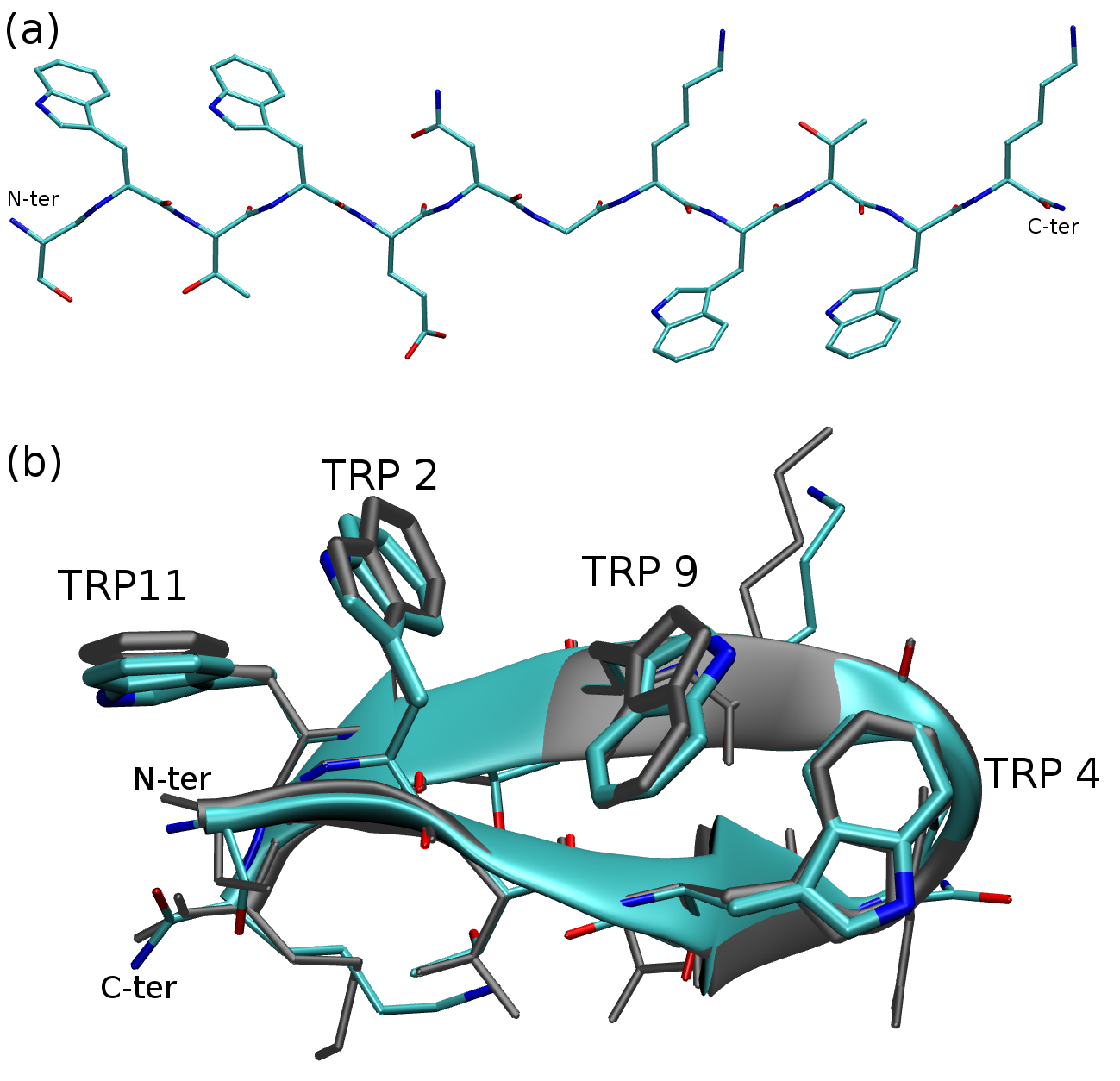}
    \end{center}
  \end{minipage}%
  \caption{\label{fig:trpzip2_best} }
\end{figure}

\begin{figure}[ht]
  \begin{minipage}{ \linewidth}
    \begin{center}
        \includegraphics[angle=90,width=\linewidth]{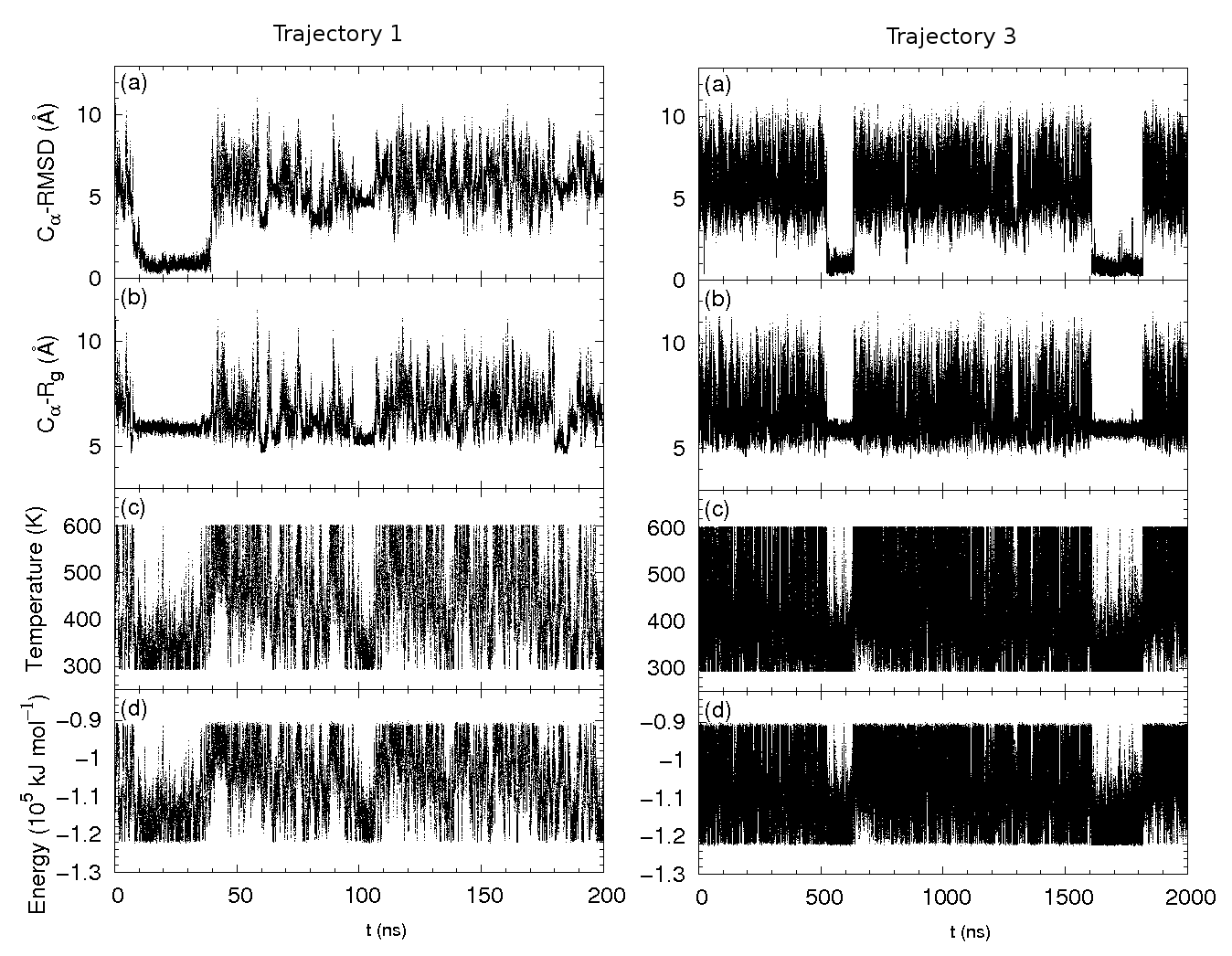}
    \end{center}
  \end{minipage}%
  \caption{\label{fig:trpzip_traj} }
\end{figure}

\begin{figure}[ht]
  \begin{minipage}{ 1.0 \linewidth}
    \begin{center}
        \includegraphics[angle=0,width=\linewidth]{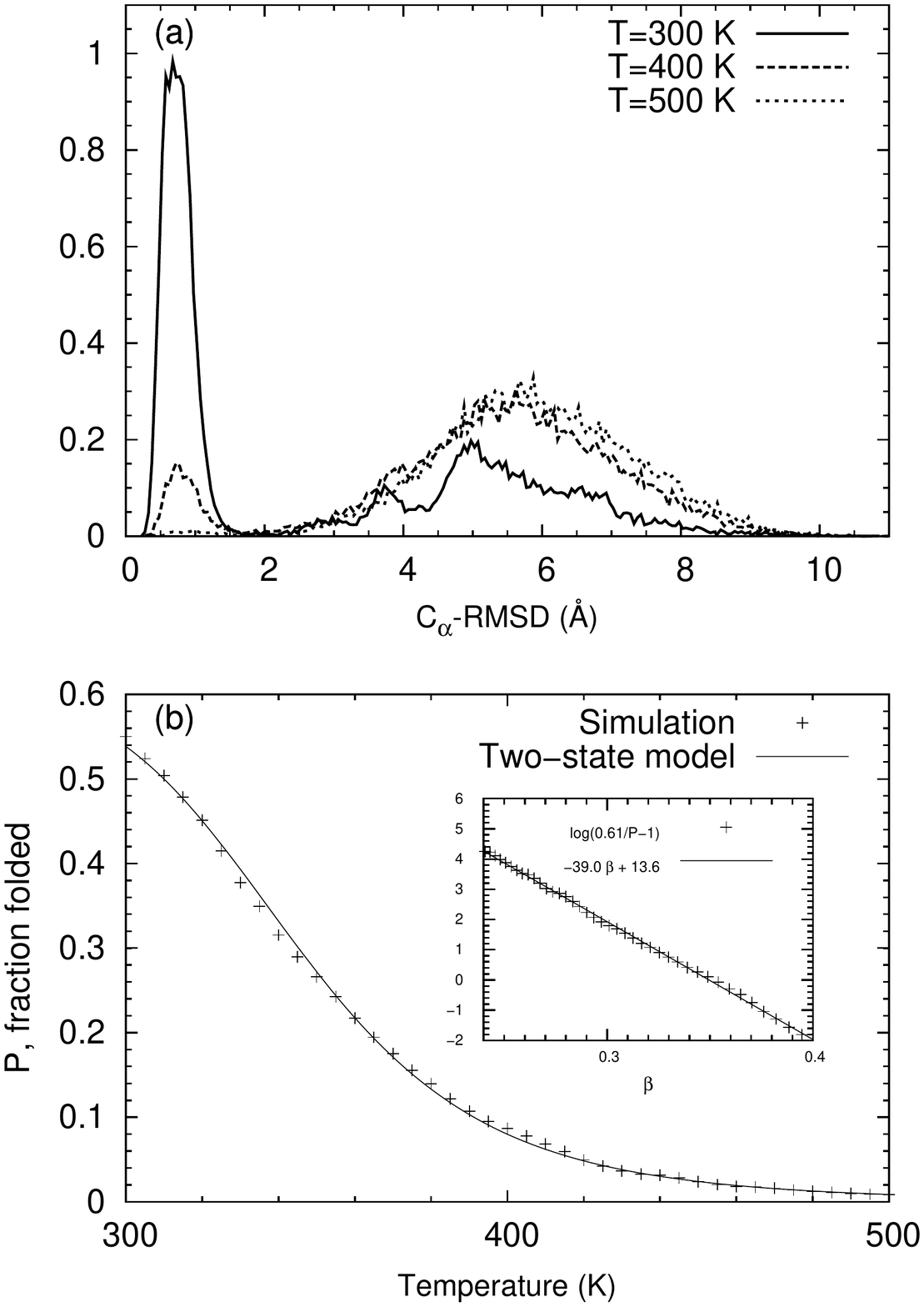}
    \end{center}
  \end{minipage}%
  \caption{\label{fig:trpzip_rhist} }
\end{figure}

\begin{figure}[ht]
  \begin{minipage}{ 1.0 \linewidth}
    \begin{center}
        \includegraphics[angle=-90,width=\linewidth]{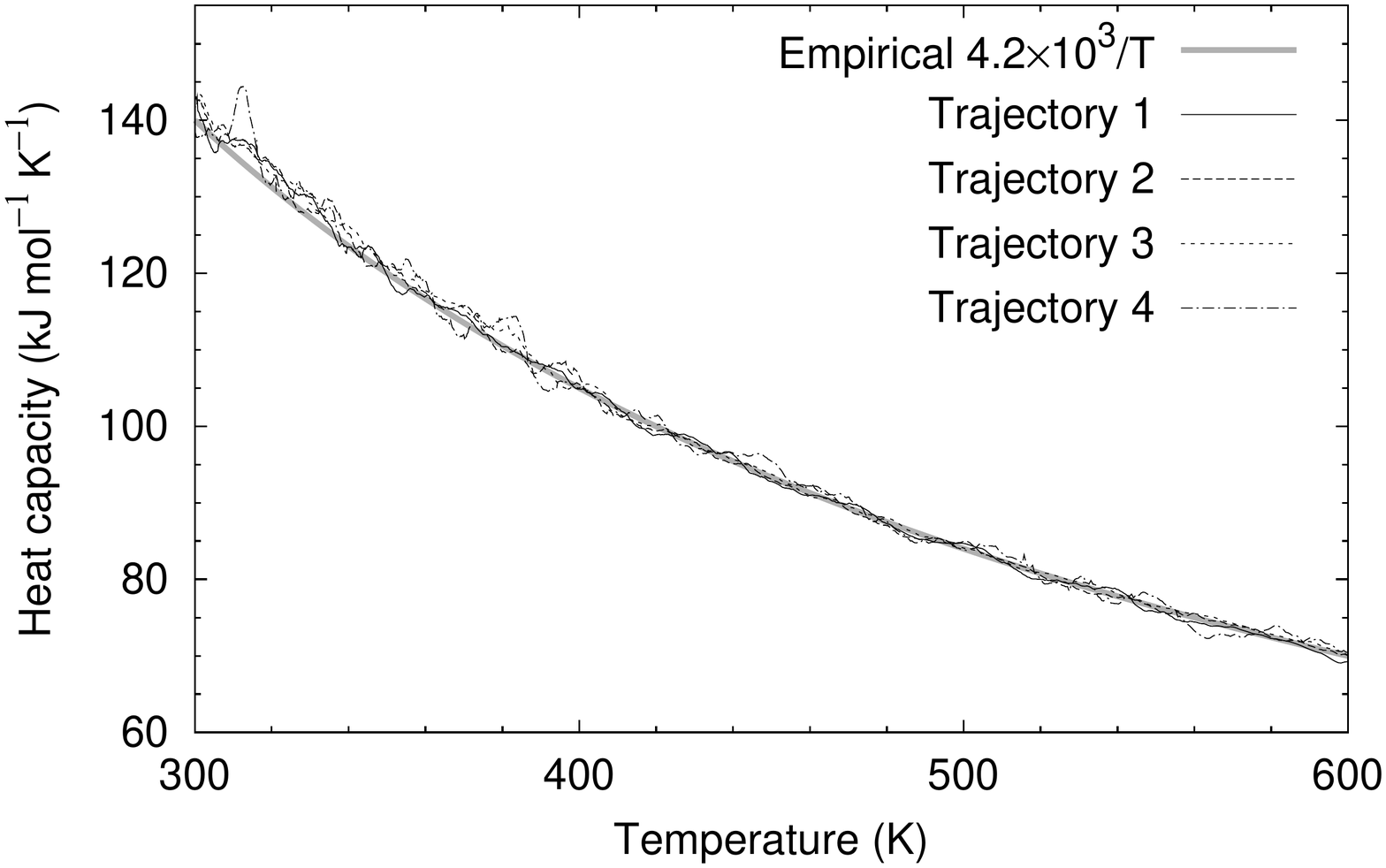}
    \end{center}
  \end{minipage}%
  \caption{\label{fig:trpzip_therm} }
\end{figure}

\begin{figure}[ht]
  \begin{minipage}{ 1.0 \linewidth}
    \begin{center}
        \includegraphics[angle=0,width=\linewidth]{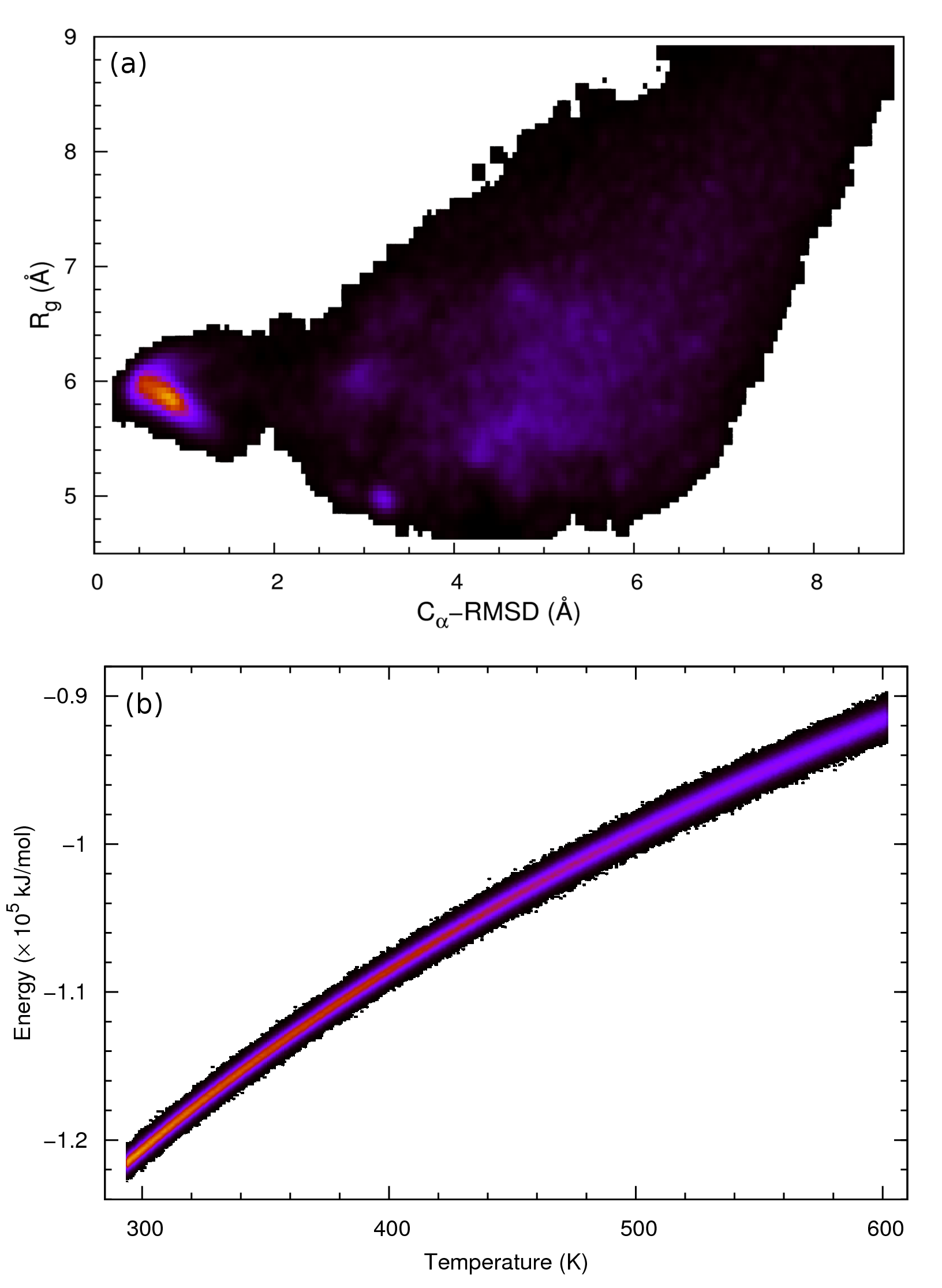}
    \end{center}
  \end{minipage}%
  \caption{\label{fig:trpzip2_grmap} }
\end{figure}


\begin{figure}[ht]
  \begin{minipage}{0.75 \linewidth}
    \begin{center}
        \includegraphics[angle=90,width=\linewidth]{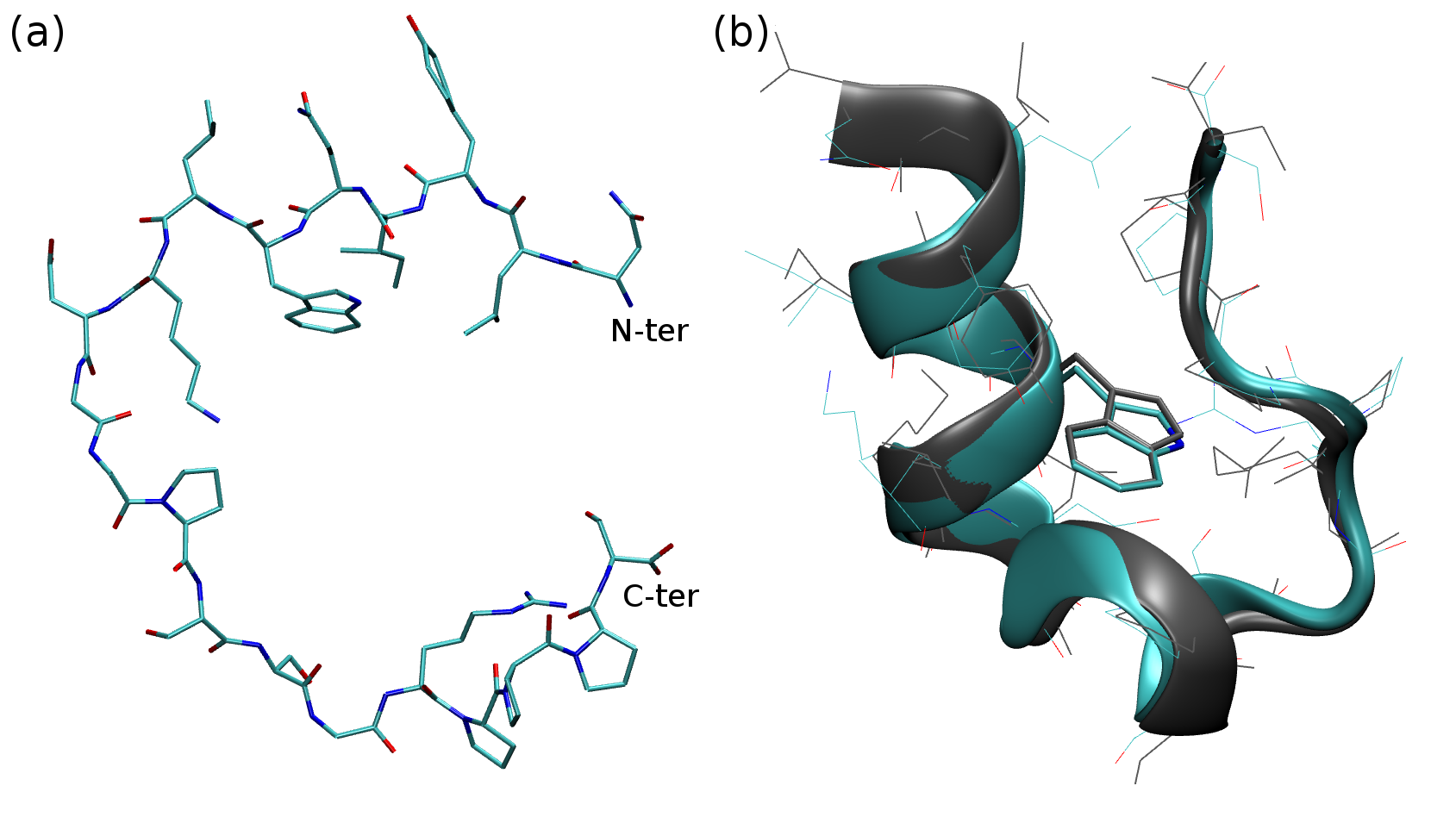}
    \end{center}
  \end{minipage}%
  \caption{\label{fig:trpcage_best} }
\end{figure}

\begin{figure}[ht]
  \begin{minipage}{0.7 \linewidth}
    \begin{center}
        \includegraphics[angle=90,width=\linewidth]{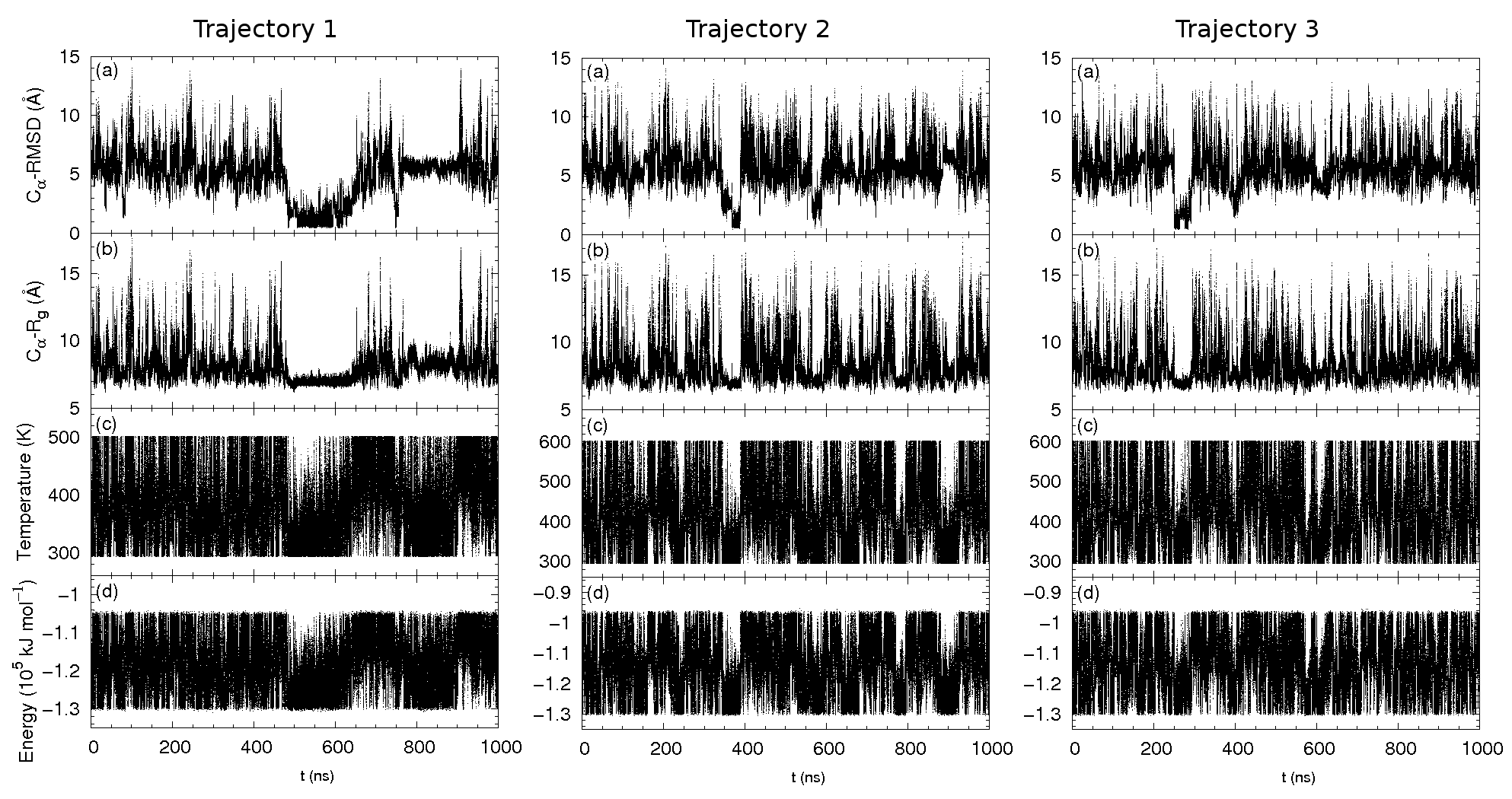}
    \end{center}
  \end{minipage}%
  \caption{\label{fig:trpcage_traj} }
\end{figure}


\begin{figure}[ht]
  \begin{minipage}{ 0.55 \linewidth}
    \begin{center}
        \includegraphics[angle=0,width=\linewidth]{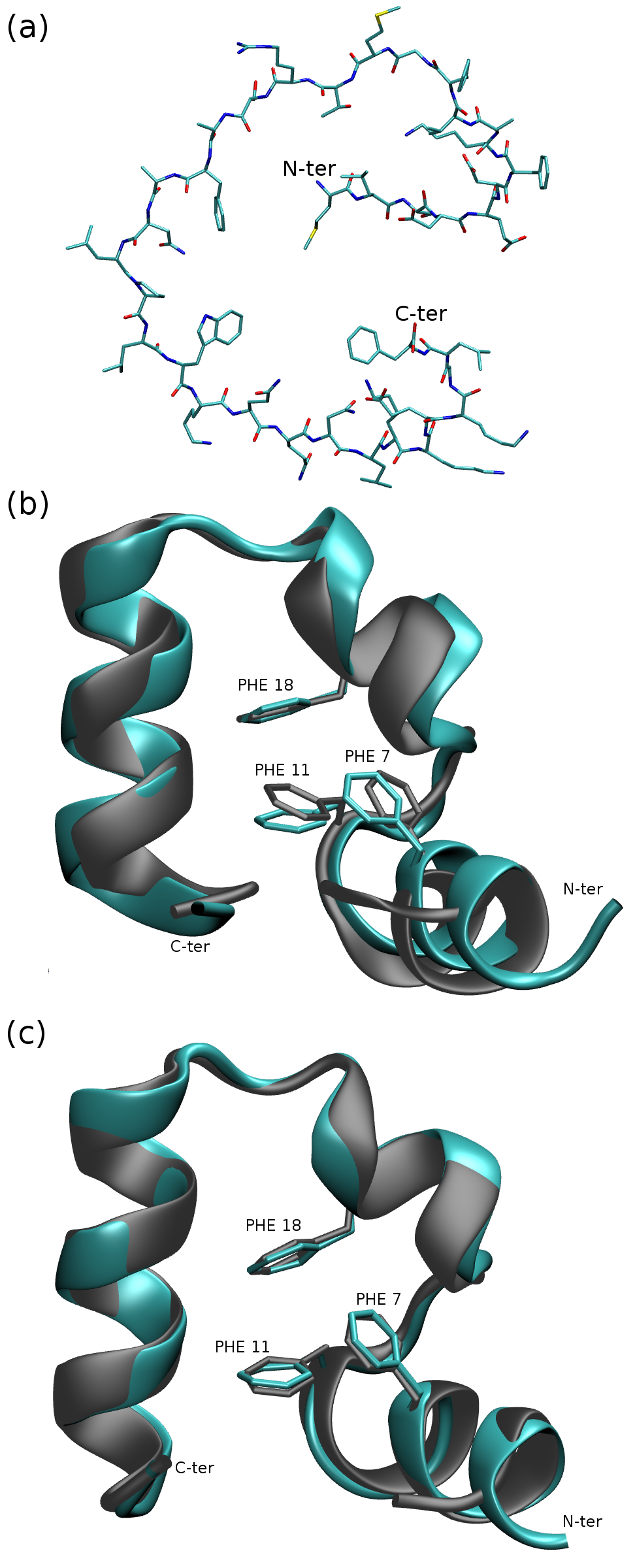}
    \end{center}
  \end{minipage}%
  \caption{\label{fig:villlin_best1} }
\end{figure}

\begin{figure}[ht]
 \begin{minipage}{ \linewidth}
   \begin{center}
       \includegraphics[angle=0,width=\linewidth]{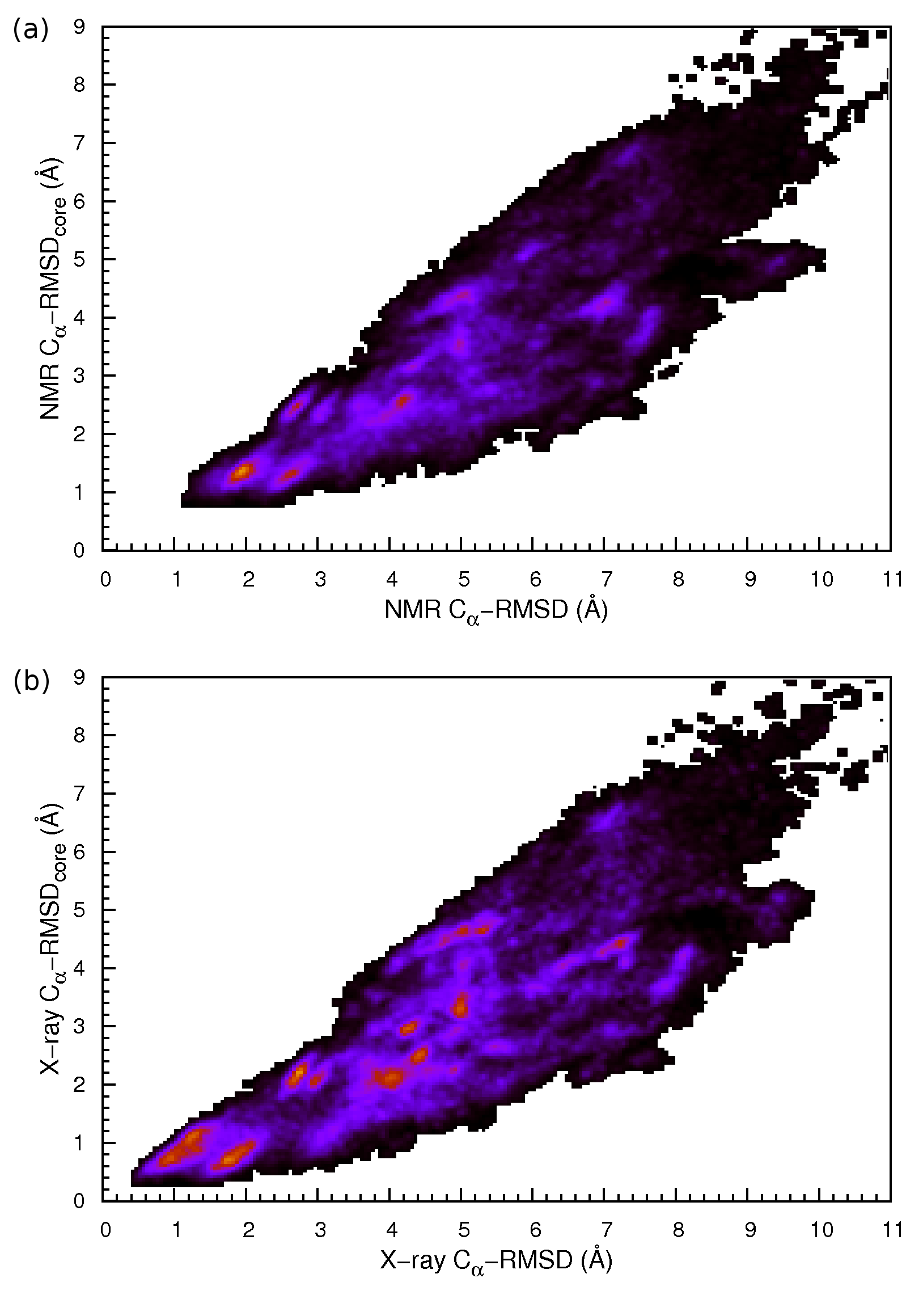}
   \end{center}
 \end{minipage}%
 \caption{\label{fig:villin_map} }
\end{figure}

\end{document}